\documentclass[11pt]{article}
\usepackage{url}
\usepackage{graphicx}
\usepackage{amsmath}
\usepackage{amssymb}

\begin{document}
\begin{center}TAGGING EMC EFFECTS AND HADRONIZATION MECHANISMS BY SEMI-INCLUSIVE DEEP
INELASTIC SCATTERING OFF NUCLEI
\vskip0.3cm

  \underline {CLAUDIO CIOFI DEGLI ATTI}$\,^1$, LEONID P. KAPTARI $\,^{1,2}$ and CHIARA BENEDETTA MEZZETTI$\,^1$\\[1ex]
\vskip0.3cm
\small{
\emph{$^1$ Department of Physics, University of Perugia, and Istituto Nazionale di Fisica Nucleare, Sezione di Perugia,
   Via A. Pascoli, I-06123, Italy}
   \vskip0.1cm
\emph{$^2$ On leave from Bogoliubov Lab. Theor. Phys., 141980, JINR, Dubna, Russia; supported through the program "Rientro dei Cervelli" of the Italian Ministry of University and Research}}
\end{center}
\vskip0.3cm
\noindent The semi-inclusive deep inelastic scattering  of electrons off a nucleus $A$  with detection of a slow nucleus $(A-1)$ in the ground or low excitation states, i.e. the process $A(e,e'(A-1))X$, can provide useful information on  the origin of the EMC effect and the mechanisms of hadronization.
The theoretical description of the process is reviewed and the results of several calculations on  few-body systems and complex nuclei are presented.
\vskip0.3cm


\vskip0.4cm
\section{Introduction}
\begin{figure}[htbp]
\begin{center}
\includegraphics[scale=0.28]{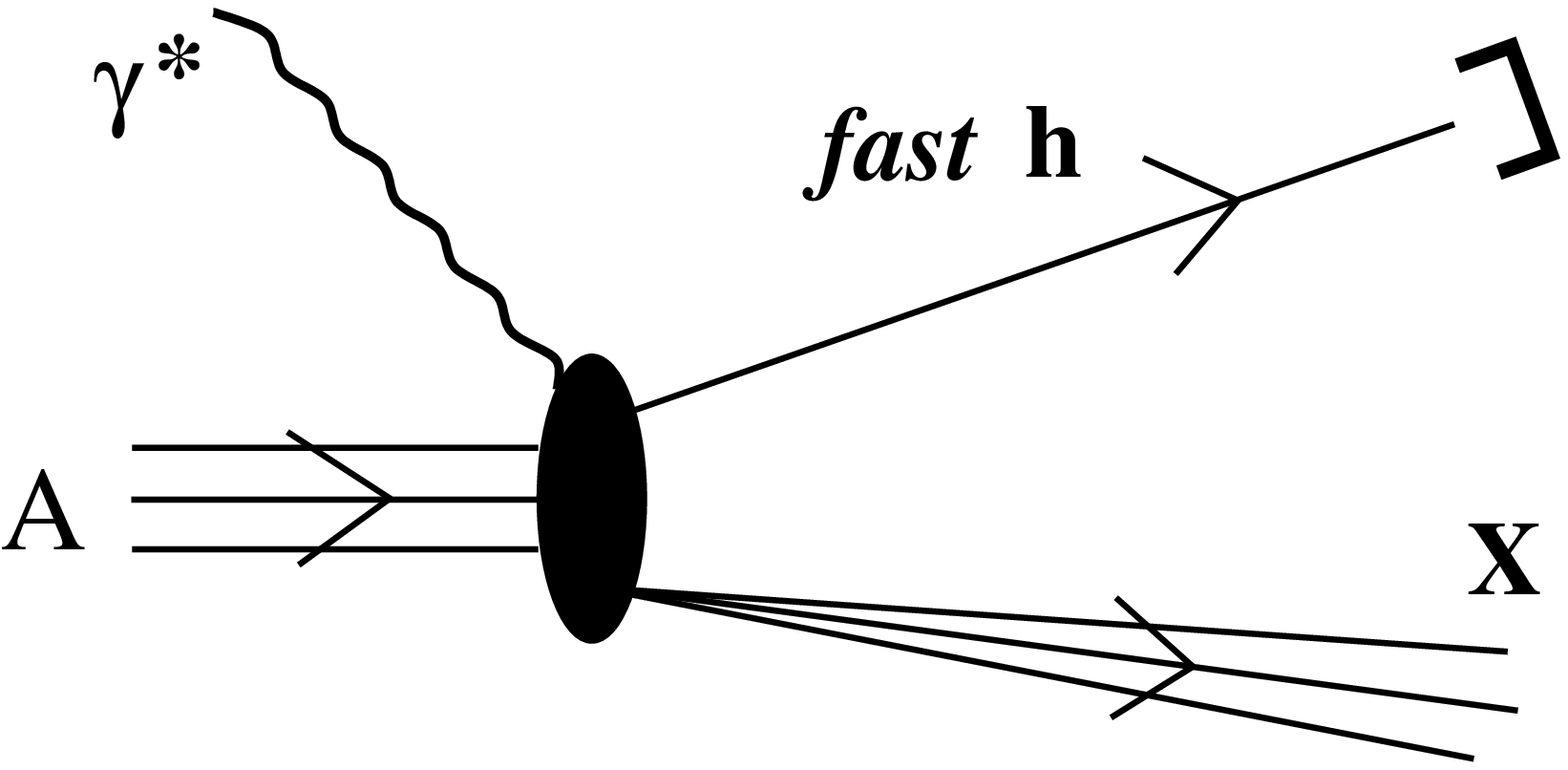}
\hskip -1.1cm
\includegraphics[scale=0.28]{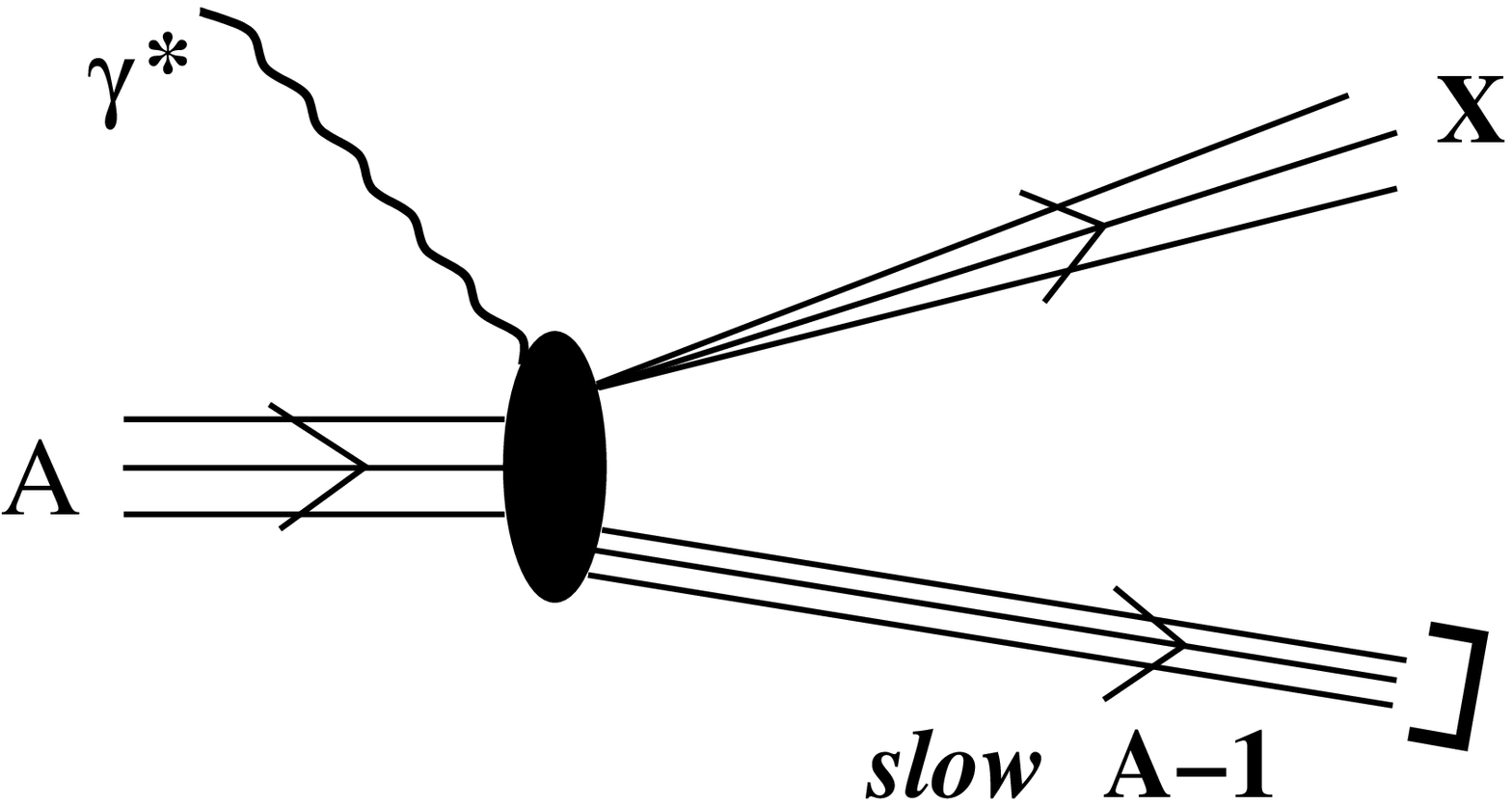}
\vskip -3.0cm
\caption{\label{Fig1} \textbf{Left}: The
''classical'' SIDIS process $A(e,e'h)X$ in one-photon exchange approximation: the fast hadron $h$,
originated from the leading quark hadronization, is detected in
coincidence with the scattered electron $e'$. \textbf{Right}:
the
new SIDIS processes $A(e,e'(A-1))X$; here a nucleus   $(A-1)$ is detected in coincidence with the scattered electron
 $e'$ .}
\end{center}
\end{figure}

\begin{figure}[htbp]
\begin{center}
\includegraphics[width=0.45\textwidth,height=0.25\textheight]{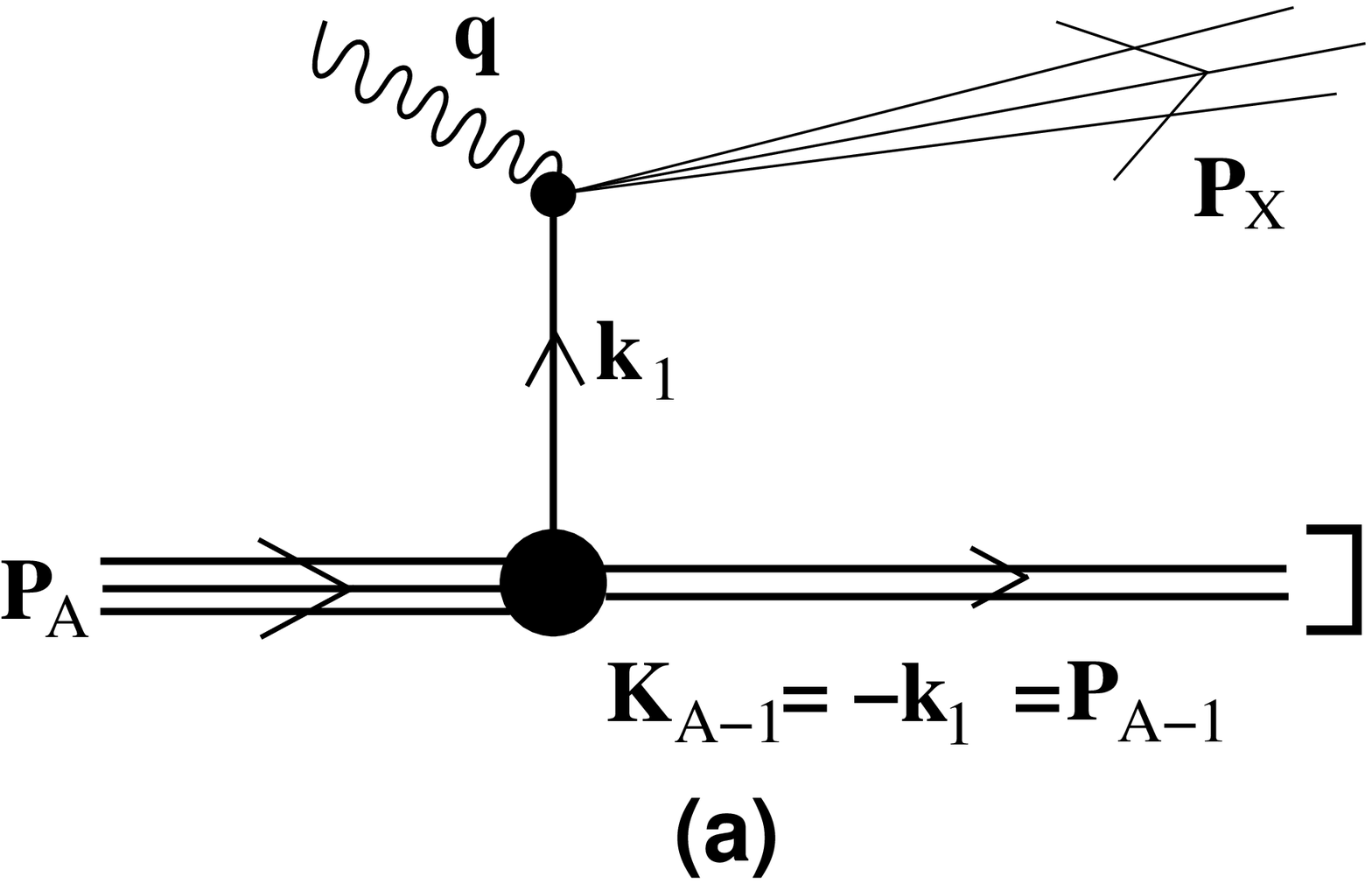}
\includegraphics[width=0.45\textwidth,height=0.25\textheight]{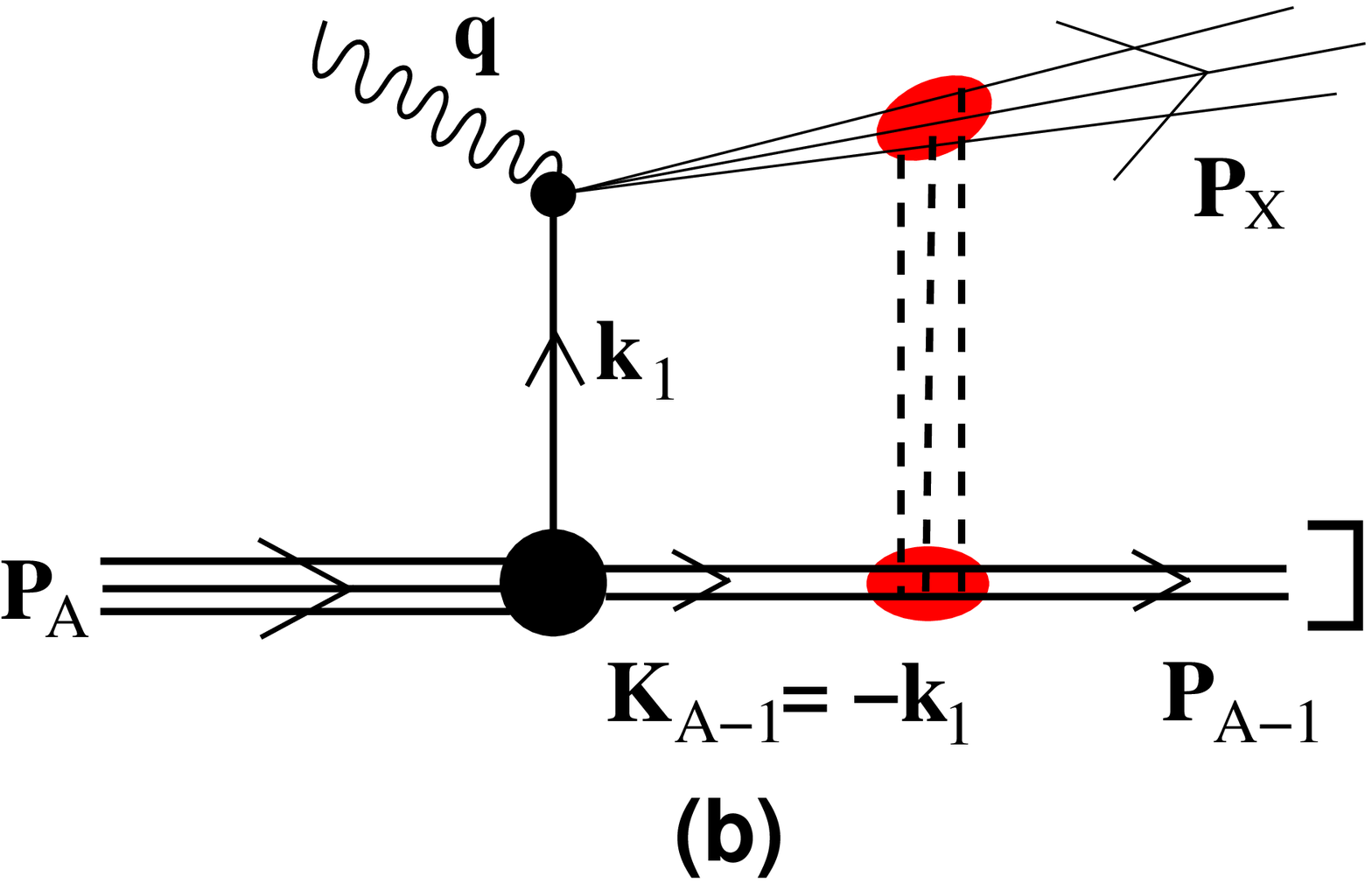}
\vskip -0.1cm
\caption{\label{Fig2} (Color online) The Feynman diagrams corresponding to the PWIA (a) and the FSI (b) cross sections
 of the SIDIS process $A(e,e'(A-1))X$ within the spectator mechanism.
 }
\end{center}
\end{figure}

In spite of many experimental and theoretical efforts (for a recent   review see \cite{EMC}),
 the origin of the nuclear EMC effect has not yet  been fully clarified and the problem as to whether and to what extent
 the quark distribution of nucleons undergoes  deformations due to the nuclear medium remains open. At the same time,
  information on hadron formation length comes, to date,  mainly from the measurement of the multiplicity ratio of the
  lepto-produced hadrons in semi inclusive  deep inelastic scattering (SIDIS) off nuclei $A(e,e'h)X$ depicted in Fig.~\ref{Fig1} ({\bf left})
  (for a recent review see \cite{brooks}).
This type of SIDIS, investigated by the HERMES experiment at HERA, as well as JLab \cite{brooks}, has provided
relevant information on different models of hadronization in the medium
\cite{koppred,ww,amp,giessen}. However, some important
details of the hadronization mechanism are still missing, e.g.  it is difficult by the SIDIS process $A(e,e'h)X$
to  obtain information on the early stage of hadronization; for such a reason, other types of SIDIS processes
 should be investigated in order to improve our knowledge on hadronization in the medium.
 To this end, the process $A(e,e'(A-1))X$,  depicted in One Photon Approximation in Fig.~\ref{Fig1} ({\bf right}),  has been proposed
  in Ref. ~\cite{scopetta} within the so called Spectator Mechanism, according to  which the virtual photon $\gamma^*$
  interacts with a quark of a nucleon of the target $A$ and the "spectator" nucleus $(A-1)$ recoils by momentum
  conservation and is detected in coincidence with the scattered electron. In Ref. ~\cite{scopetta}, however,
  the Plane Wave Impulse Approximation (PWIA) (diagram \ref{Fig2}(a)) has been used, so that some of the results were
  only of qualitative character.
\noindent
An important step forward in the investigation of the new SIDIS process was made in Ref.~\cite{ciokop}, where the theoretical
approach has been   extended by  considering   the final state interaction (FSI) of the hadronizing debris (the leading quark and the diquark)
with the nucleons of the nucleus $(A-1)$ (diagram \ref{Fig2}(b)) in terms of an effective debris-nucleon  interaction cross sections
derived on
the basis of the hadronization model of Ref.~\cite{koppred}.
Once the propagation and interaction of the debris with the spectators are taken into account, the advantages of the SIDIS
  process $A(e,e'(A-1))X$
become clear.
Concerning the  hadronization mechanism, it has been shown \cite{ciokop} that the survival probability of the recoil
nucleus $(A-1)$ is very sensitive to the details of the rescattering and hadronization of the debris, i.e. to the mechanism  of the
hadronization.
Moreover, a clear advantage of the process under discussion, compared to leading hadron production, is the
possibility to study the early stage of hadronization at short formation times without being affected by cascading
processes. Indeed, no cascading is possible if the recoiling  nucleus $(A-1)$ survives. At the same time,  most of hadrons
with small momentum produced in inclusive process $A(e,e'h)X$ originate from cascading of more energetic particles, so that, in order to analyze data and
extract information on the early stage of hadronization,  a
realistic model for cascading is necessary, which  is barely possible  \cite{ciokop}. As for the investigation of the EMC
effects, it has been shown \cite{scopetta} that a properly defined  ratio of the cross section on nucleus A,
taken at a value of the Bjorken scaling variable $x_{Bj}$, to the cross section on the same nucleus,
taken at a different value of $x_{Bj}$, is essentially identical to the ratio of the nucleon structure
functions $F^{(N/A)}(x_{Bj})/F^{(N/A)}(x_{Bj}^{\prime})$.

The  SIDIS process $A(e,e'(A-1))X$ was almost impossible to access experimentally when it was proposed
in $1999$, but notable progress has been done since then:  in a recent experiment at Jefferson Lab \cite{klimenko,kuhn},
the process
$^2H(e,e'p)X$ has been investigated in detail  finding, as  will be shown later on, a gratifying agreement with theoretical
results \cite{ckk,ck2}; moreover,  experimental proposals have been finalized to study the process on complex nuclei,
thanks to the development of proper recoil detectors \cite{proposals}.

The aim of this paper is to review the theoretical description of the process $A(e,e'(A-1))X$ and to present the results of
several calculations for both few nucleon systems and complex nuclei.

The SIDIS cross section which includes FSI  has the following form \cite{ckk,ck2,veronica}
\begin{eqnarray}
   &&\!\!\!\!\!\! \sigma^{A,FSI} (x_{Bj},Q^2,|\textbf{P}_{A-1}|,y_A,z_1^{(A)})
   \equiv\sigma^{A,FSI}=
  \frac{d\sigma^{A,FSI}}{d x_{Bj} d Q^2  d  \textbf{ P}_{A-1}}=\nonumber\\&&
   =  K^A( x_{Bj},Q^2,y_A,z_1^{(A)}) z_1^{(A)}
   F_2^{N/A}(x_A,Q^2,k_1^2)\, n_0^{A,FSI}( \textbf{P}_{A-1}),
   \label{crossdist}
   \end{eqnarray}
where
 \begin{equation}
  K^A( x_{Bj},Q^2,y_A,z_1^{(A)})=
  \frac{4\alpha_{em}^2}{Q^4} \frac {\pi}{x_{Bj}}\cdot \left( \frac{y}{y_A}\right)^2
  \left[\frac{y_{A}^2}{2} + (1-y_A) -
  \frac{k_1^2x_{Bj}^2 y_A^2}{z_1^{(A)2}Q^2}\right]
  \label{ka}
  \end{equation}
with
\begin{eqnarray}  y_A = \frac{k_1\cdot q}{k_1\cdot
k_e} ~, \quad
 x_A = {x_{Bj} \over z_1^{(A)}}, \quad
z_1^{(A)} = {k_1 \cdot q \over m_N \nu}~.
\label{adef}
\end{eqnarray}
Here $k_1$ is the four-momentum of the bound nucleon, $q^2 =
(E_e-E_{e'})^2 - ({\bf k}_e-{\bf k}_{e'})^2= -Q^2$ and,
eventually,   $n_0^{A,FSI}(\textbf{P}_{A-1})$ is the  distorted
momentum distribution of the bound nucleon
\begin{eqnarray}
\hspace*{-1cm} &&n_0^{A,FSI}(\textbf{P}_{A-1}) = \frac{1}{2J_A+1}\nonumber\\
\hspace*{-1cm}&&\sum_{{\cal M}_A,{\cal M}_{A-1}} \left | \int\, d {\bf r}_1^{\prime}
 e^{i {\bf P}_{A-1} {\bf r}_1^{\prime}} \langle \Psi_{J_{A-1}, {\cal M}_{A-1}}^{0}
( \{{\bf r}_i^{\prime}\}) |S_{FSI}^{XN}|
\Psi_{J_{A}, {\cal M}_{A}}^{0}( {\bf r}_1^{\prime}, \{{\bf r}_i^{\prime} \}) \rangle
 \right |^2\!\!\!\!.
 \label{dismomfsi}
 \end{eqnarray}
\noindent In Eq. (\ref{dismomfsi}), the primed quantities denote intrinsic coordinates and the quantity
$S_{FSI}^{XN}$ is the debris-nucleon eikonal scattering
$S$-matrix
 \begin{equation}
S_{FSI}^{XN}\equiv S_{FSI}^{XN}({\bf r}_1,\dots,{\bf r}_A)=
\prod_{i=2}^{A}\bigl[1-\theta(z_i-z_1)\Gamma({\bf b}_1-{\bf b}_i,{
z}_1-{z}_i)\bigr] \label{SG}
\end{equation}
where  the $Q^2$- and $x_{Bj}$-dependent profile function is
\begin{equation} \Gamma^{XN}({{\bf b}_{1i}},z_{1i})\,
=\,\frac{\sigma_{eff}(z_{1i}, Q^2,x_{Bj})}
{4\,\pi\,b_0^2}\,\exp \left[-\frac{{\bf
b}_{1i}^{2}}{2\,b_0^2}\right],
 \label{eikonal}
\end{equation}
with ${\bf r}_{1i}=\{{\bf b}_{1i},z_{1i}\}$ being  $z_{1i} =z_{1}-z_{i}$
and ${\bf b}_{1i}={\bf b}_{1}-{\bf b}_{i}$. It can be seen that, unlike the standard Glauber eikonal
approach ~\cite{glau2}, the profile function $\Gamma^{XN}$ depends not only upon the  transverse relative
 separation but also upon the longitudinal separation $z_{1,i}={z}_1-{z}_i$ due to the $z$- (or time) dependence
 of the effective cross section $ \sigma_{eff}(z_{1i}, Q^2,x_{Bj}) \equiv \sigma_{eff}(z_{1i})$.
The cross section (\ref{crossdist}) describes the process in which, after the hard interaction of $\gamma^*$ with a
quark of the so called ''active'' nucleon ''N'', a nucleon debris is created, composed by a nucleon $N_1$ arising from
target fragmentation, and a color string which propagates and hadronizes giving rise to an increasing with time (distance)
number of pions. The interaction of the hadronizing debris with the nucleons of $(A-1)$ is described by an effective cross
 section $\sigma_{eff}(z)$, which depends on the total energy of the debris, $W_X^2\equiv P_X^2$; if the energy
  is not high enough, the hadronization procedure can terminate inside the nucleus $(A-1)$, after which  the number of
  produced hadrons and the  cross section $\sigma_{eff}(z)$ remain constant.
The effective debris-nucleon cross section derived in \cite{ciokop} has the following form
\begin{equation}
\sigma_{eff}(z)\,=\,\sigma_{tot}^{NN}\,+\,\sigma_{tot}^{\pi N}\,\big[\,n_{M}(z)\,+\,n_{G}(z)\,\big]
\label{sigmaeff}
\end{equation}
where the $Q^2$- and $x_{Bj}$-dependent quantities $n_{M}(z)$ and $n_{G}(z)$  are the pion
 multiplicities due to the breaking of the color string and to gluon radiation, respectively, whose explicit forms
 are given in Ref. \cite{ciokop}.

\section{Comparison between experimental data and theoretical calculations for the process $^2H(e,e'p)X$}
Experimental data on the process $^2H(e,e'p)X$ have  recently been
obtained at Jlab~\cite{klimenko,kuhn} in the following kinematical
regions: beam energy $E_e = 5.75\,\, GeV$, four-momentum transfer
$1.2 \,\,(GeV/c)^2 \lesssim Q^2 \lesssim 5.0 \,\,(GeV/c)^2$,
recoiling proton momentum $0.28 \,\,GeV/c \lesssim|{\bf
p}_{p}|\leq 0.7\,\, GeV/c$, proton emission angle $-0.8  \leq
\cos\theta_{\bf p} \leq 0.7$ $(\theta_{\widehat {{\bf
p}_p\cdot{\bf q}}} \equiv \theta_{\bf p})$, invariant mass of the
produced hadronic state $1.1 \,\,GeV\,\,\leq W_X\leq \,\,2.7 \,\,
GeV$, with $W_X^2={(k_1+q)}^2={(P_D- p_p+q)}^2$ (in what follows
all deuteron quantities: cross sections, mass, momentum, etc. will
be denoted by a case $D$). The experimental data have been plotted
in terms of the reduced cross section
\begin{figure}[htbp]
\begin{center}
 \includegraphics[width=0.47\textwidth,height=0.35\textheight]{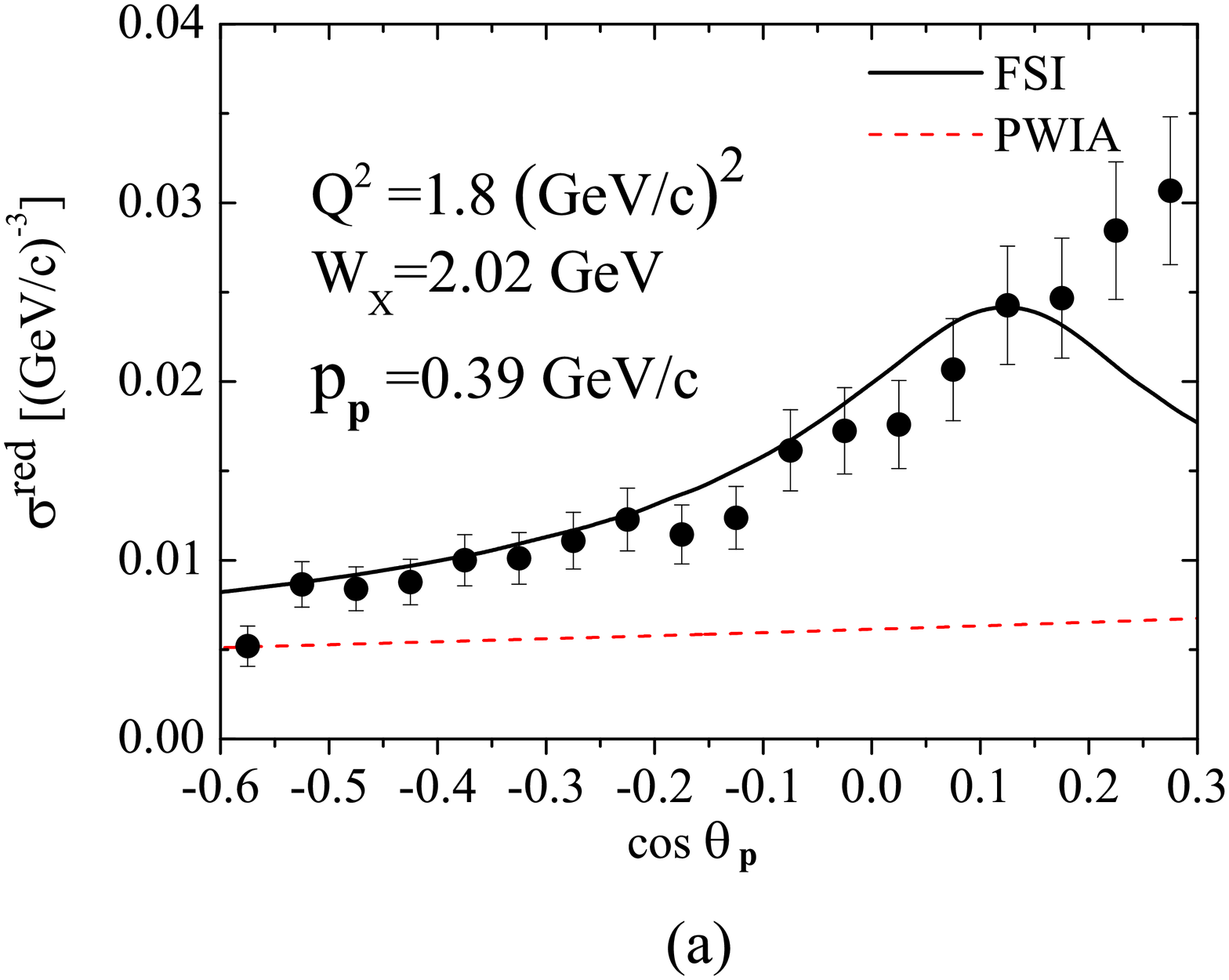}
 \includegraphics[width=0.47\textwidth,height=0.35\textheight]{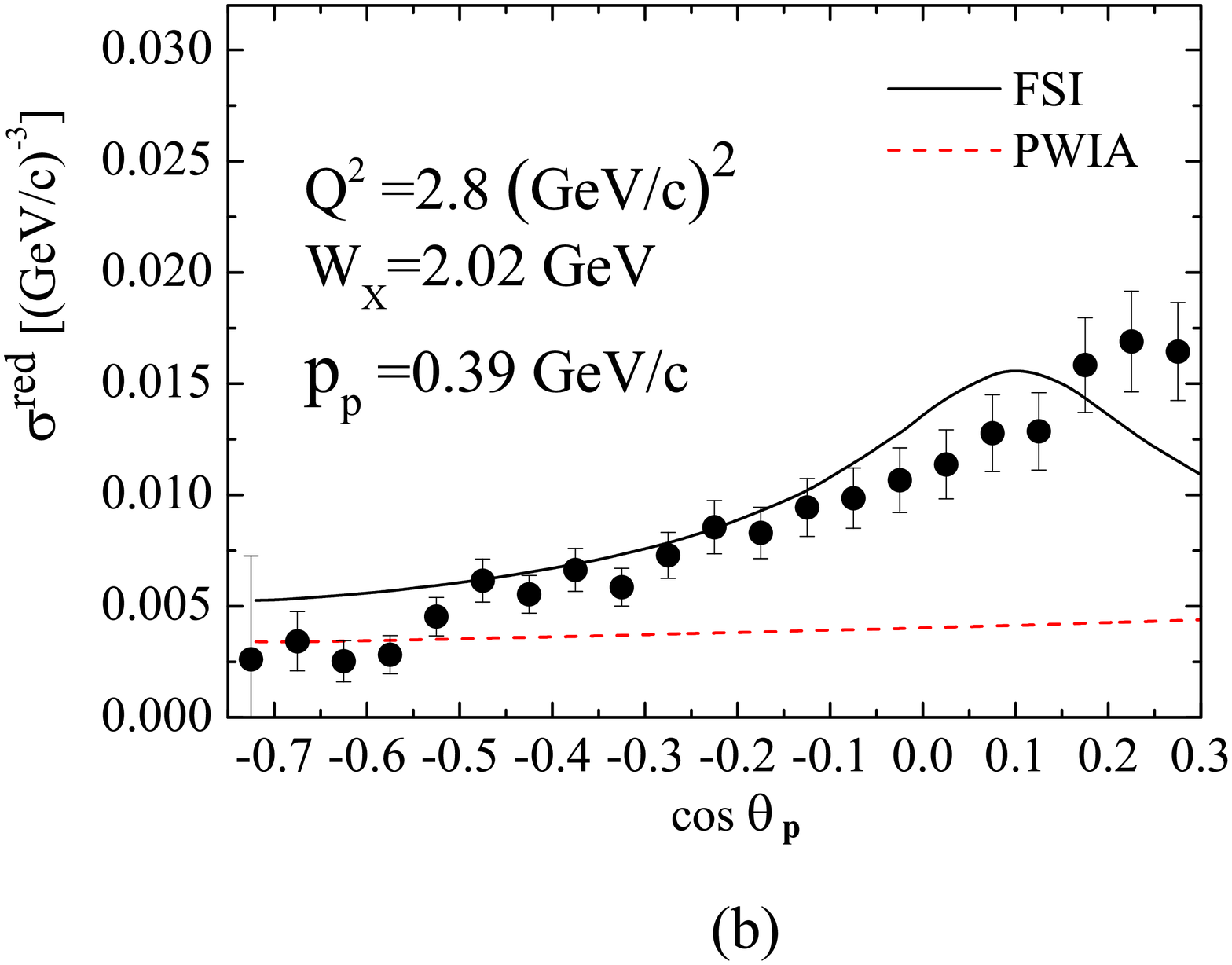}\vspace*{-10mm}
 \includegraphics[width=0.47\textwidth,height=0.35\textheight]{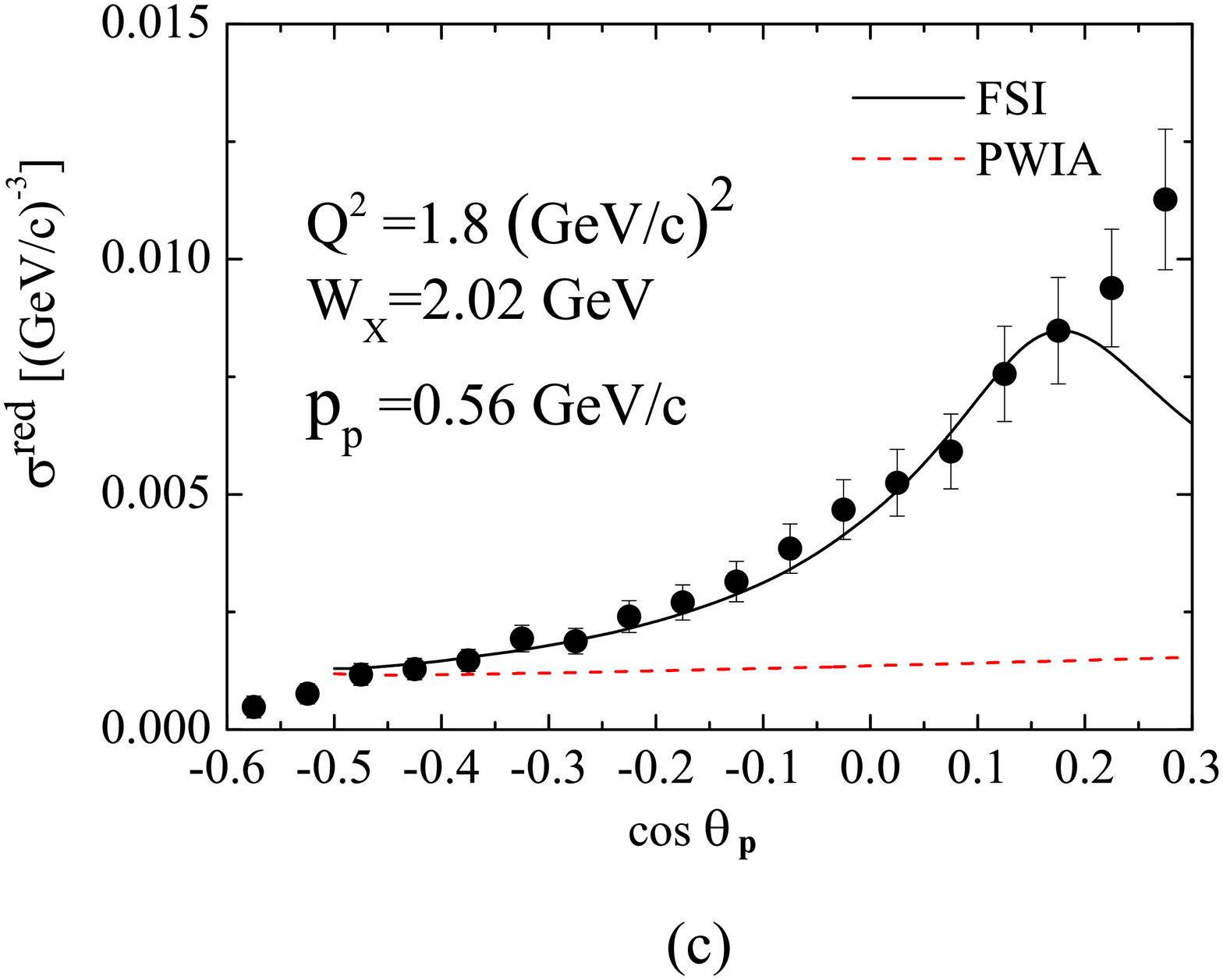}
 \includegraphics[width=0.47\textwidth,height=0.35\textheight]{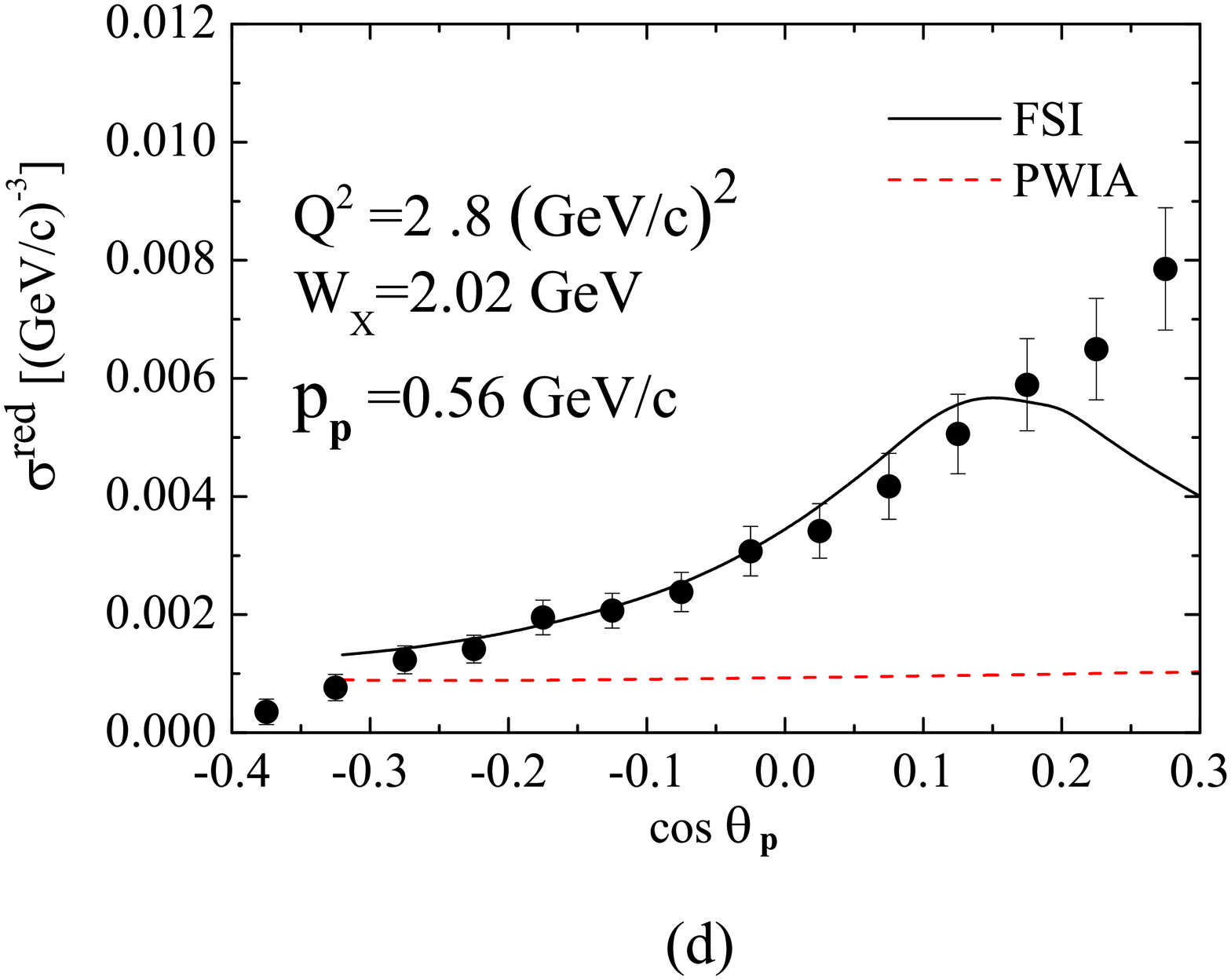}
\vskip -0.5cm
\caption{\label{Fig3} (Color online)The theoretical reduced cross section, Eq. (\ref{reducedour}), {\it vs}.
$cos \theta_{\widehat{{\bf q}\cdot {\bf p}_p}}$  ($\theta_{\widehat{{\bf q} \cdot {\bf p}_p}} \equiv \theta_p$)
 compared with the experimental data of Refs. \cite{klimenko, kuhn}.
 Each Figure shows the reduced cross section calculated at fixed values of the four momentum transfer $Q^2$,
 the invariant mass  $W_X$ of the hadronic state $X$, and the momentum $|{\bf p}_p| \equiv p_p$ of the detected proton.
 The error bars represent the sum in quadrature of statistical and systematic errors given in
  Refs. \cite{klimenko,kuhn} (after Ref. \cite{ck2}).}
\end{center}
\end{figure}
\begin{equation}
\sigma^{red}(x_{Bj},Q^2,{\bf p}_p) = \frac{1}{K^A(
x_{Bj},Q^2,y_A,z_1^{(A)}) }\,\left(
\frac{y}{y_D}\right)^2\frac{d\sigma ^{D,exp}}{d x_{Bj} d Q^2  d
\textbf{p}_p} \label{reduced}
\end{equation}
which, considering Eq. (\ref{crossdist})
would be
 \begin{equation}
\sigma^{red}(x_{Bj},Q^2,{\bf p}_p) = \left( \frac{y}{y_D}\right)^2
z_1^{(D)} F_2^{N/D}(x_D,Q^2,k_1^2)\, n_0^{D,FSI}({\bf p}_{p})
\label{reducedour} \end{equation}
\noindent in agreement with the experimental definition  of Ref. \cite{klimenko}.
A comparison between theoretical calculations (performed using the AV18 potential \cite{AV18})
 and the experimental data plotted {\it vs} $\cos \theta_{{\bf p}}$
at fixed values of $Q^2$, $W_X$ and  $|{\bf p}_p|$, is presented in Fig.~\ref{Fig3}, which clearly shows that:
i) apart from the very backward emission, the experimental data are dominated by the FSI;
ii) the  model of FSI of Refs. \cite{ciokop,ckk,ck2} provides a satisfactory description of the experimental data in
the backward direction and also around $\theta_{{\bf p}} \simeq 90^{\circ}$ (a comparison of theoretical
 results and experimental data in the full range of kinematics of  Refs.~\cite{klimenko, kuhn} will be
  presented elsewhere; iii) in the forward direction ($\theta_{{\bf p}} \lesssim 80^{\circ}$) the spectator mechanism
   fails to reproduce the data and it is clear that other production mechanisms are playing a role in this region.
   For such a reason in what follows only   the region ($\theta_{{\bf p}} \gtrsim 80^{\circ}$) will be considered,  where useful
   information on both the hadronization mechanism and the EMC effect can in principle be obtained, provided the
   data are analyzed in the proper way, i.e. getting rid of  EMC effects, in the former case, and of nuclear effects,
   in the latter case.
\section{The process $A(e,e'\,(A-1))X$ on few body systems and complex nuclei}

\begin{figure}[htbp]
\begin{center}
\includegraphics[scale=0.35]{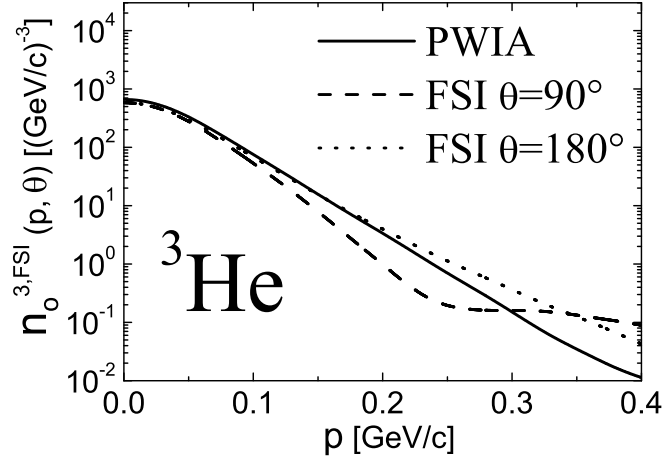}
\vskip 1cm
\includegraphics[scale=0.35]{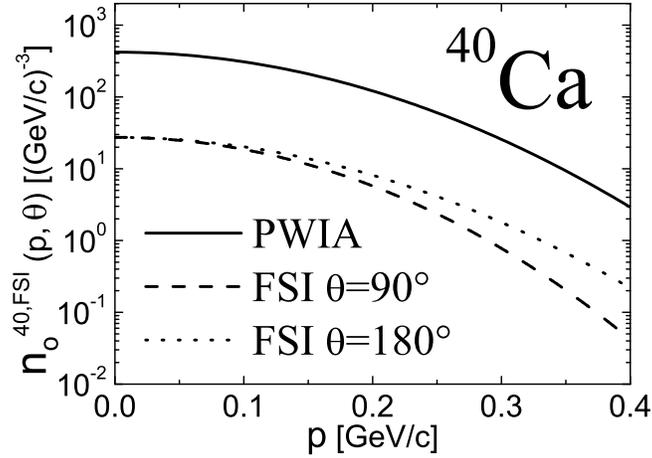}
\vskip -0.5cm
\caption{\label{Fig4} The distorted momentum distributions
$n_0^{A,FSI}({\bf P}_{A-1})$ (Eq. (\ref{dismomfsi}))  for $^3He$
 and $^{40}Ca$ . The full curves  correspond to the
PWIA ($\sigma_{eff}=0$) and the dotted and dashed curves to the
FSI ($\sigma_{eff} \neq 0$) in parallel ($\theta_{\widehat{{\bf q}
{\bf p}_p}} \equiv \theta=180^{0}$) and perpendicular
($\theta_{\widehat{{\bf q} {\bf p}_p}}\equiv \theta =90^{0}$)
kinematics, respectively.}
\end{center}
\end{figure}

\begin{figure}[htbp]
\begin{center}
\includegraphics[scale=0.45]{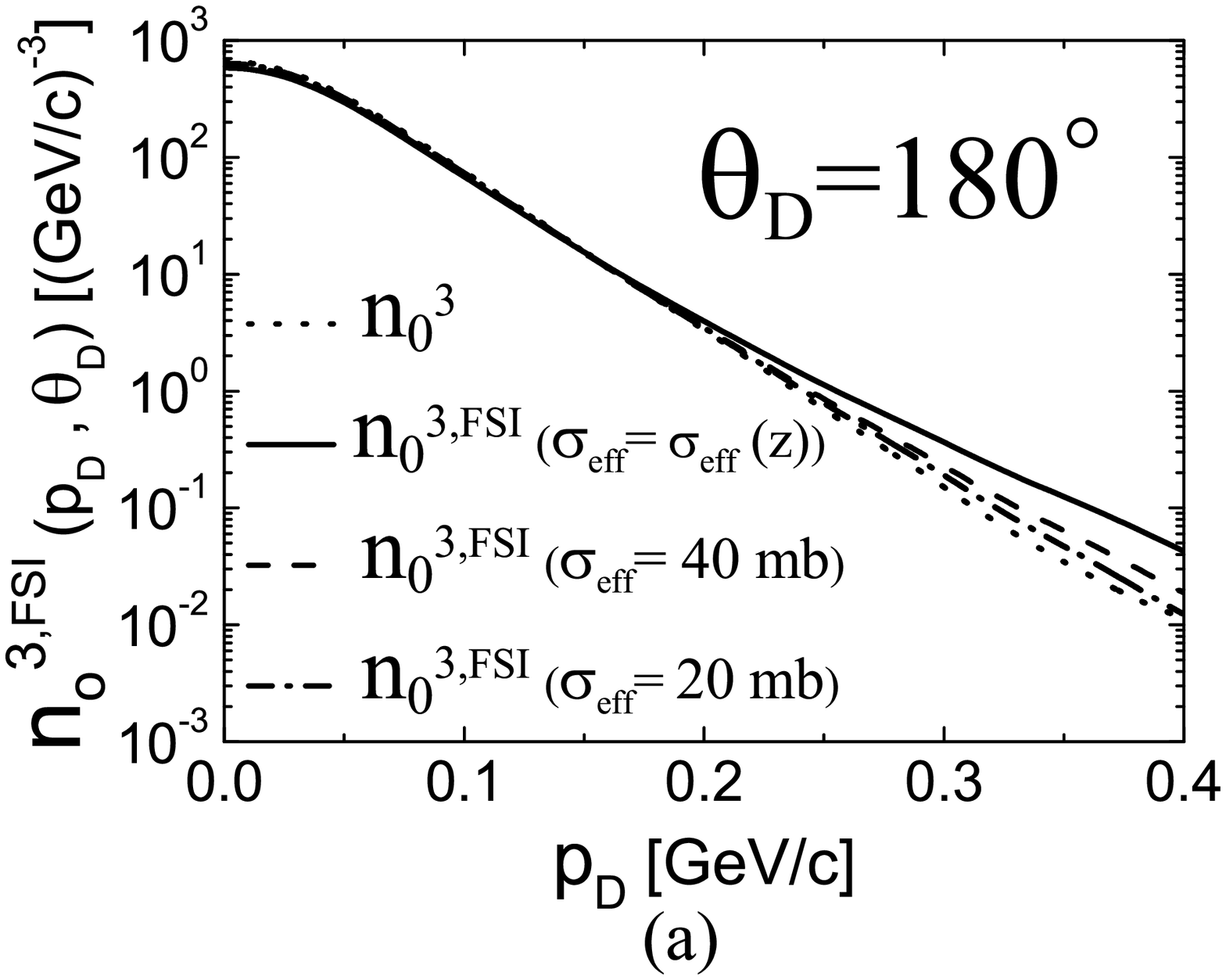}
\vskip-1.0cm
\includegraphics[scale=0.45]{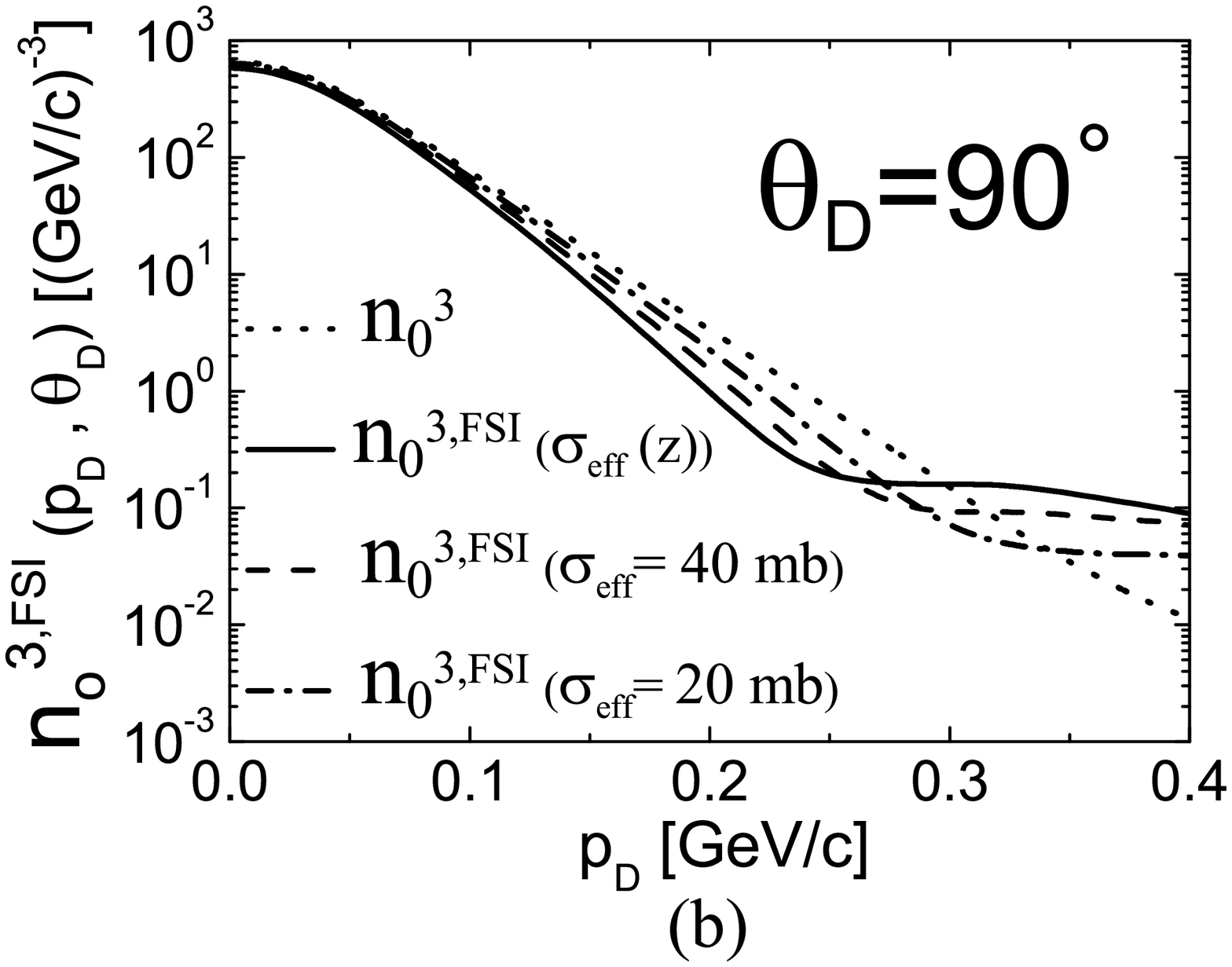}
\vskip -0.5cm
\caption{\label{Fig5} The distorted momentum distributions $n_0^{3,FSI}({\bf P}_{A-1})$ (Eq. (\ref{dismomfsi})
with ${\bf P}_{A-1} \equiv {\bf p}_{D}$) in the process $^3He(e,e'd)X$ in  parallel
($\theta_{\widehat{{\textbf{P}_{A-1}\cdot \textbf{q}}}}\equiv \theta_D=180^{0}$) ({\bf a})
and  perpendicular ($\theta_{\widehat{{\textbf{P}_{A-1}\cdot \textbf{q}}}}\equiv \theta_D=90^{0}$) ({\bf b}) kinematics calculated with
different effective debris-nucleon cross sections in Eq. (\ref{eikonal}): the effective  debris-nucleon
cross section $\sigma_{eff}(z) \equiv \sigma_{eff}(z,Q^2,x_{Bj})$  (full line) and two constant cross sections (dashed and dot-dashed lines).
The undistorted momentum distribution $n_0^{3}(|{\bf P}_{A-1}|)$ is shown by the dotted line. Calculations have been performed
at the following kinematics:  $E_e=12 \ GeV$, $Q^2=6 \ GeV^2/c^2$ and $W_X^2=5.8\
GeV^2$ (after Ref. \cite{ck2}).}
\end{center}
\end{figure}

We have considered  the processes $^3He(e,e'd)X$ and $^{40}Ca(e,e'^{39}K)X$. Calculations have been performed using for $^2H$
and $^3He$ \cite{rosati} wave functions
corresponding   to a realistic interaction \cite{AV18}, whereas for heavier nuclei single
particle mean field wave functions have been adopted.
The form of the nucleon structure function $F_2^{N/A}$ is from Ref.~\cite{effe2},   with the nucleon
off-mass shell  within the x-rescaling model, i.e. $x_A=x_{Bj}/z_1^{(A)}$ where   $z_1^{(A)}=k_1\cdot
q/(m_N\nu)$,  with  $k_1^0 =M_A-\sqrt{(M_{A-1}^*)^2+{{\bf K}_{A-1}}^2}$.
In Fig. \ref{Fig4}, the $^3He$ and $^{40}Ca$ distorted momentum distributions given by
 Eq. (\ref{dismomfsi}) are shown
 at parallel ($\theta_{\widehat{{\textbf{P}_{A-1}\cdot \textbf{q}}}}\equiv \theta=180^{0}$) and perpendicular
 ($\theta_{\widehat{{\textbf{P}_{A-1}\cdot \textbf{q}}}}\equiv \theta=90^{0}$) kinematics. The sensitivity of the
 distorted momentum distributions upon different models of the debris-nucleon cross section is illustrated
 in  Fig. \ref{Fig5}.

\begin{figure}[htcp]
\begin{center}
\includegraphics[scale=0.5]{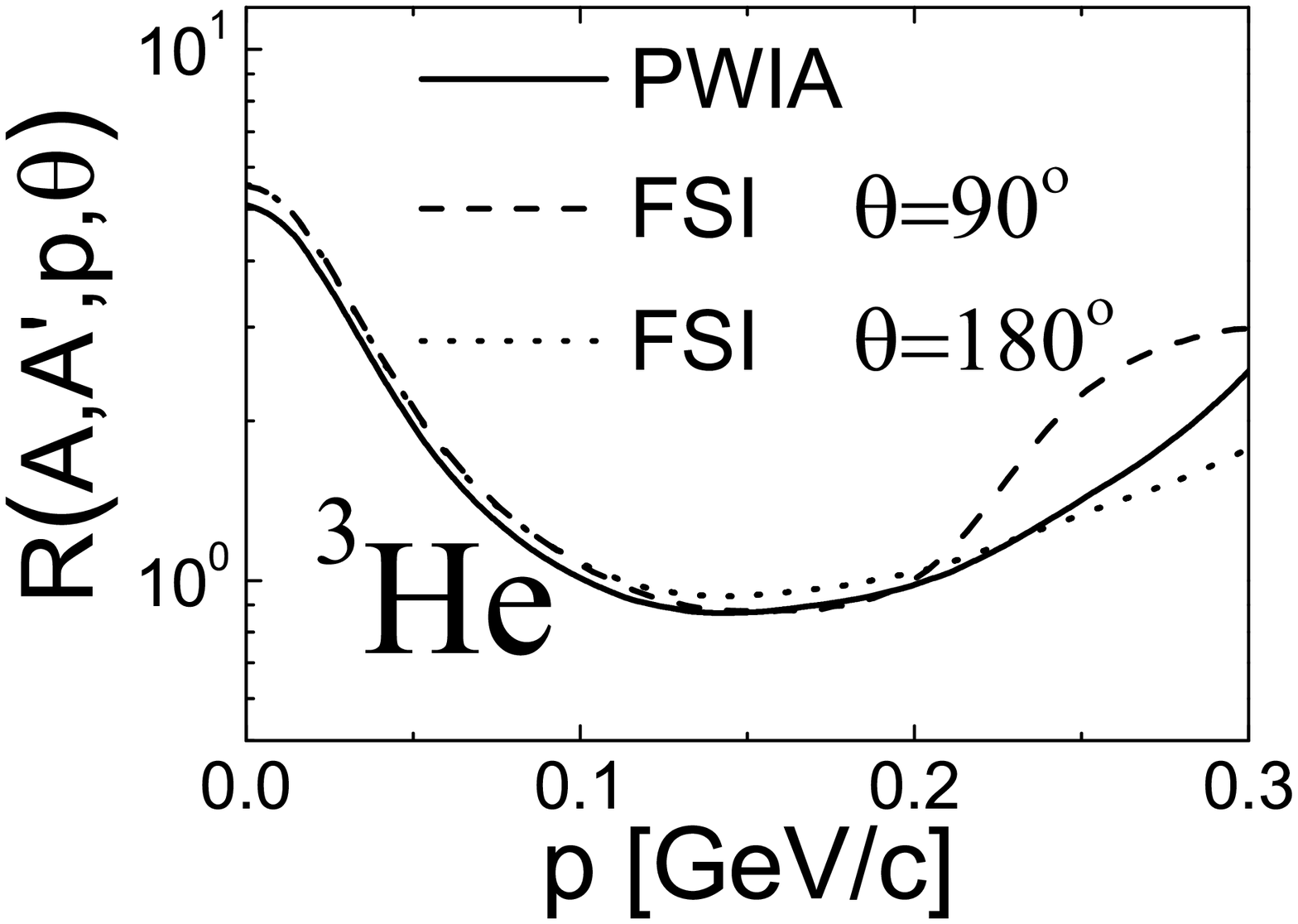}
\includegraphics[scale=0.5]{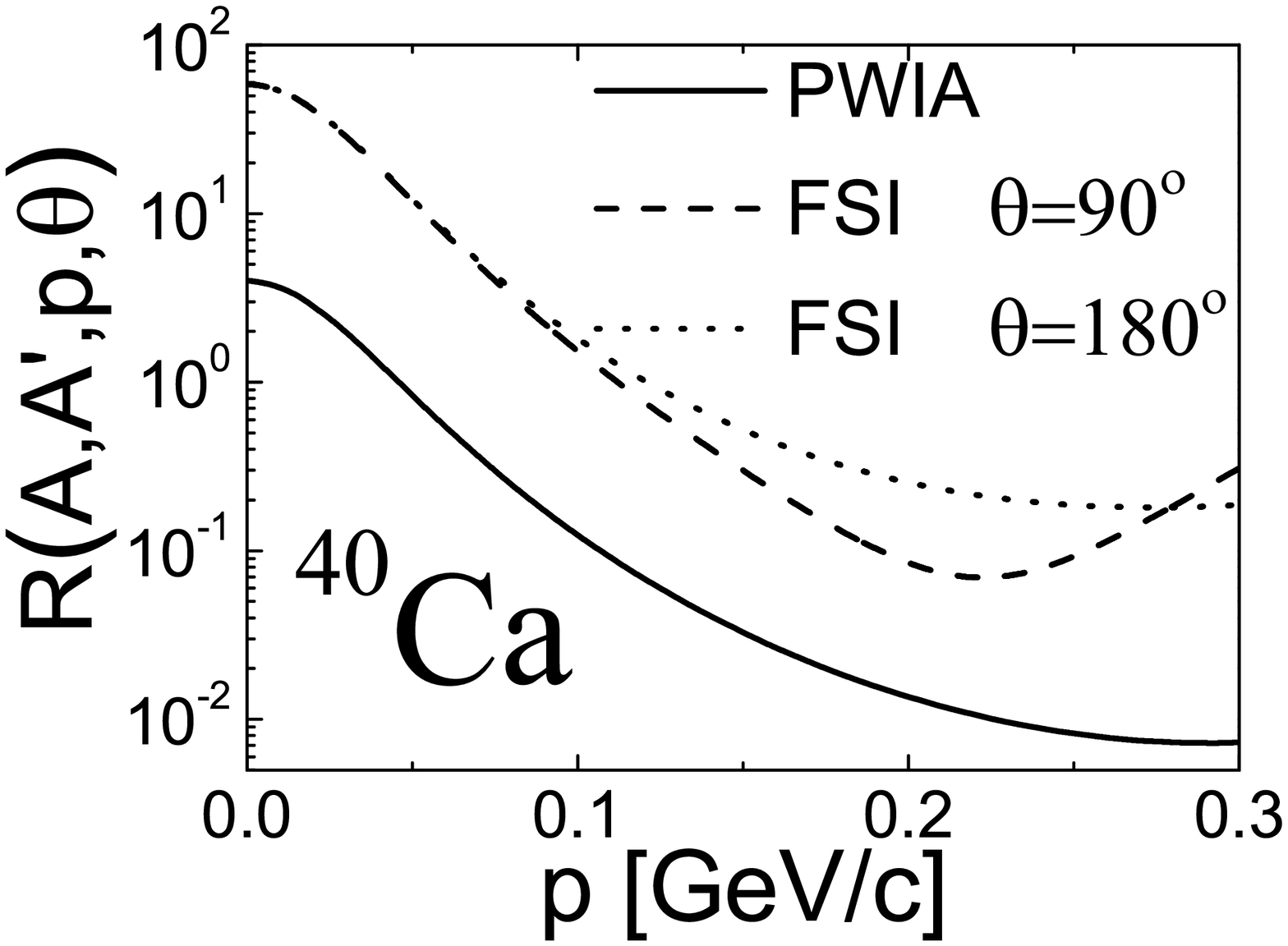}
\vskip-0.5cm \caption{\label{Fig6} The ratio (\ref{ratioa-1})
with  $A =2$, $A'=3$ (\textbf{left}) and $A =2$, $A'=40$
(\textbf{right}). \emph{Full curve}: PWIA ($\sigma_{eff}=0$);
\emph{dashed and  dotted curves}: FSI ($\sigma_{eff} \neq 0$) in
perpendicular
 and  parallel
  kinematics, respectively ($p\equiv |\textbf{p}| \equiv |\textbf{P}_{A-1}|$).}
\end{center}
\end{figure}

It can be seen that, in the case of few-nucleon systems, FSI is particularly relevant at high momentum values,
whereas for complex nuclei the momentum distributions are strongly affected also at low momentum, with a resulting
strong decrease of the survival probability of $(A-1)$ (see Ref. \cite{ciokop} for more details).
In order to obtain information on hadronization mechanisms minimizing, at the same time,  possible contaminations
from the poor knowledge of the neutron structure function, the ratio of the cross sections for
two different nuclei $A$ and $A^{\prime}$, measured at the same value of $x_{Bj}$, should be considered \cite{scopetta,ck2}, {\it viz}
 \begin{equation}
\frac{\sigma^{A,exp} ( x_{Bj},Q^2,|\textbf{P}_{A-1}|,z_1^{(A)},y_A ) }
 {\sigma^{A',exp} ( x_{Bj},Q^2,|\textbf{P}_{A-1}|,z_1^{(A')},y_{A'} ) }
 \rightarrow \frac{{n_0^{(A,FSI)}(\textbf{P}_{A-1})}}{{n_0^{(A',FSI)} (\textbf{P}_{A-1})}} \equiv R(A,A',\textbf{P}_{A-1})
  \label{ratioa-1}
  \end{equation}
where the last step is strictly valid only in the Bjorken limit.
However, as discussed in detail in Refs. \cite{scopetta,ck2} the ratio (\ref{ratioa-1}) is  governed
  by the ratio of the distorted momentum distributions, and any reasonably expected  A-dependence
  of $F_2^{N/A}(x_A,Q^2,k_1^2)$, through $x_A$, will not affect it.
The ratio (\ref{ratioa-1}) for $A=2$, $A'=3$ and $A'=40$ is shown in Fig.~\ref{Fig6}.
Since the low momentum part of the momentum distributions is very well known, the experimental
observation of strong deviations from the PWIA predictions (full curves) would provide information on
the debris-nucleon interaction and, consequently, on hadronization. Experiments on  heavier nuclei, particularly at perpendicular
 kinematics and $|{\bf P}_{A-1}| \simeq 0.2 \div 0.3 \,\,GeV/c$ ({\it cf.} Fig. \ref{Fig6}),  where the effects of
 FSI are expected to be more relevant \cite{ciokop}, would be extremely useful to clarify the  hadronization mechanisms.

\begin{figure}[htcp]
\begin{center}
\includegraphics[scale=0.45]{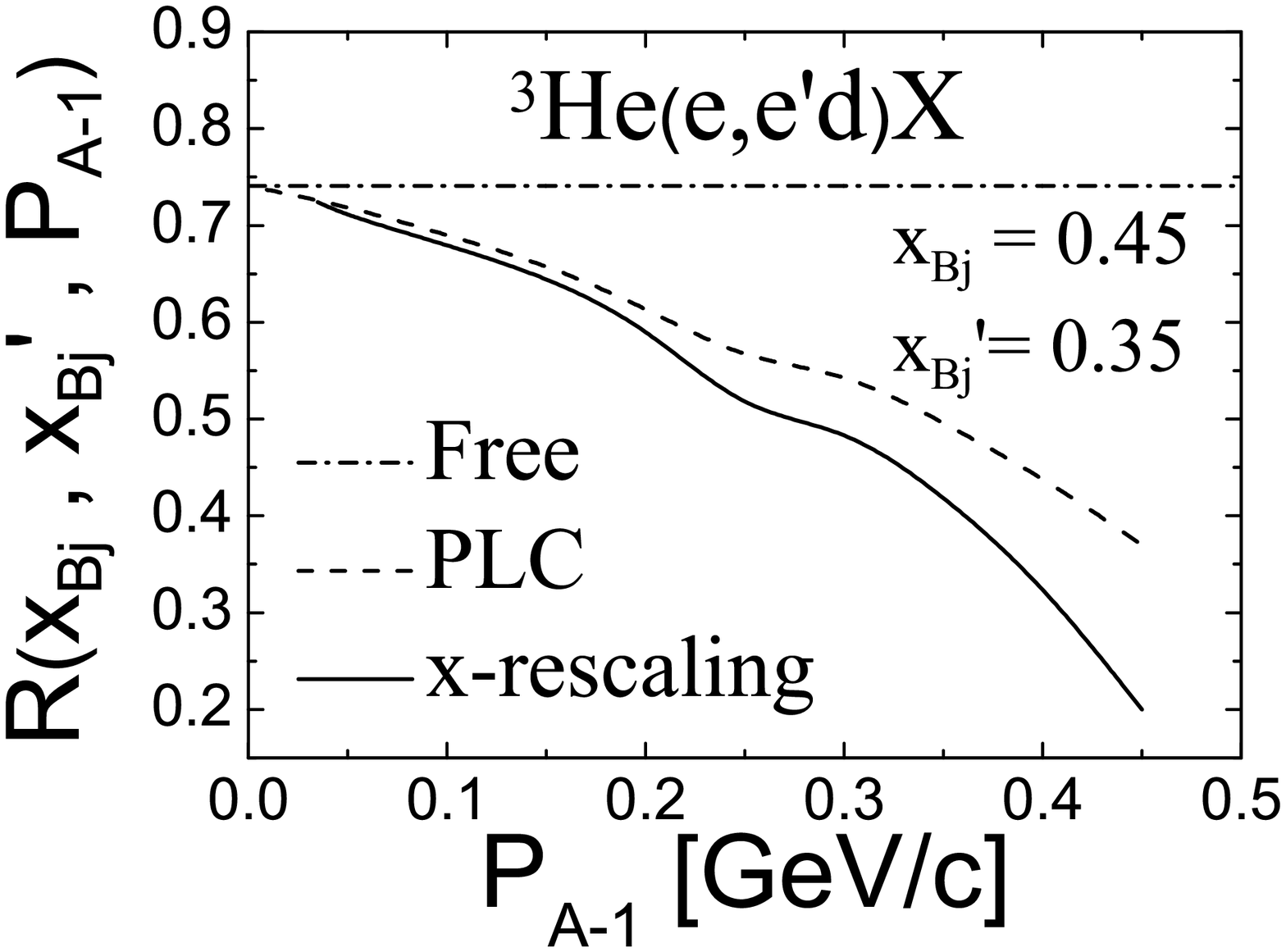}
\vskip-1.0cm
\includegraphics[scale=0.45]{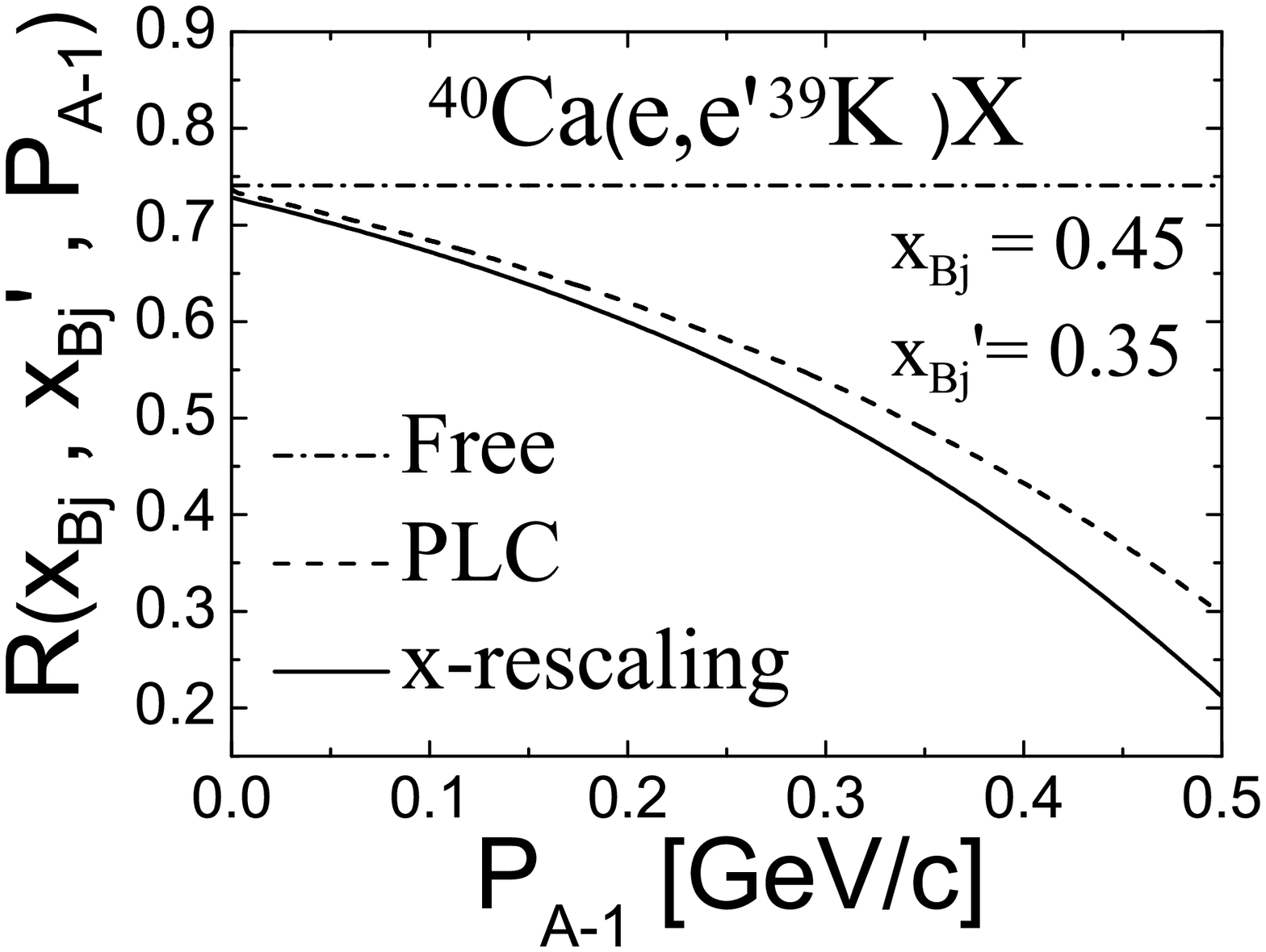}
\vskip-0.5cm \caption{\label{Fig7}
 The ratio  (\ref{ratioa}) corresponding to
the process $^3He(e,e'd)X$  and to the process
$^{40}Ca(e,e'\,^{39}K)X$  calculated at two values
of the Bjorken scaling variable $x_{Bj}$ and with different
nucleon structure functions: i)
 Free structure function (dot-dashed line): $F_2^{N/A} \left ( { x^A}  \right )
=F_2^{N/A} \left ( { x_{Bj}}  \right )$; ii) off mass-shell
(x-rescaling) structure function (full line): $F_2^{N/A} \left ( { x^{A}}\right ) =F_2^{N/A} \left
( { x_{Bj}/z_1^A}  \right )$ with $z_1^A= k_1 \cdot q/(m_N \,\nu)$ and $k_1^{0} = M_A -
\sqrt{{(M_{A-1}^{*})}^2+{\bf P}_{A-1}^2}$; iii)
 structure function with reduction of point-like configurations (PLC) in the medium depending upon
 the nucleon virtuality  $v({\bf k}_1,E)$ (Eq. (\ref{virtuality})) \cite{CKFS} (dashed line):
$F_2^{N/A} \left ( { x^{A}} \right )
=F_2^{N} \left ( { x_{Bj}/z_1^N}  \right )\cdot \delta_A(x_{Bj},v({\bf k}_1,E))$ with
$z_1^A= k_1 \cdot q/(m_N \,\nu)$ and   $k_1^{0} = M_A -
\sqrt{m_N^2+{\bf P}_{A-1}^2}$.}
\end{center}
\end{figure}

In order to tag the EMC effect, i.e. if, how,   and to what extent the nucleon structure function in the medium differs from the
free structure function, one has to get rid  of  the  distorted nucleon momentum distributions
and other nuclear structure effects by considering a quantity which would depend only upon $F_2^{N/A}(x_{A},Q^2,k_1^2)$.
This can be achieved by introducing the ratio of the cross sections on  the nucleus $A$ measured at two different values of the Bjorken scaling variable $x_{Bj}$ and $x_{Bj}^{\prime}$, leaving unchanged all other quantities in the two cross sections,  i.e. the ratio
\begin{eqnarray}
&&  \frac{\sigma^{A,exp} ( x_{Bj},Q^2,|\textbf{P}_{A-1}|,z_1^{(A)},y_A ) }
             {\sigma^{A,exp} (  x_{Bj}^{\prime},Q^2,|\textbf{P}_{A-1}|,z_1^{(A)},y_{A} ) }\nonumber \\
             &\rightarrow& \frac{F_2^{N/A}(x_A,Q^2,k_1^2)}
{F_2^{N/A}(x_{A}^{\prime},Q^2,k_1^2)}  \equiv
   R(x_{Bj}, x_{Bj}^{\prime}, |\textbf{P}_{A-1}|).
   \label{ratioa}
\end{eqnarray}
The quantity  (\ref{ratioa}) for the processes $^3He(e,e'd)X$ and $^{40}Ca(e,e'\,^{39}K)X$ were calculated in the
following kinematical range: $2\,\, GeV{^2}\,\lesssim W_X^2 \lesssim \,\, 10\,\,GeV{^2}$ and $Q^2= 8\, (GeV/c)^2$.
At each  value of $W_X$, we have changed  $|{\bf P}_{A-1}|$ from zero to  $|{\bf P}_{A-1}| = 0.5 \,\,GeV/c$,
 obtaining  for different values of $|{\bf P}_{A-1}|$  different values of $x_{Bj}$. In order to minimize the
  effects of FSI, the  angle $\theta_{\widehat {{\bf P}_{A-1}\cdot{\bf q}}} \equiv \theta_{A-1}$ has been chosen  in  the  backward
   direction, $\theta_{\widehat {{\bf P}_{A-1}\cdot{\bf q}}} \equiv \theta_{A-1}\sim 145^o $ ({\it cf.} Fig.~\ref{Fig7}).
Within such a kinematics the effective cross section $\sigma_{eff}(z_{1i},x_{Bj},Q^2)$
is the same for different values of $W_X$  and,  correspondingly,
the  distorted momentum distributions $n_0^{A,FSI}$ will depend only upon  $|{\bf P}_{A-1}|$
and cancel in the ratio (\ref{ratioa}).
By this way, all nuclear structure effects, except possible effects of in-medium deformations
of the nucleon structure function $F_2^{N/A}$, are eliminated, and one is left with a ratio which depends
only upon the nucleon structure function $F_2^{N/A}$.
Calculations  have been performed using three  different  structure functions
$F_2^{N/A}(x_{A},Q^2,k_1^2)$, namely:
\begin{enumerate}
\item
the free nucleon structure function from Ref.~\cite{effe2}, exhibiting no EMC effects;
\item
the nucleon structure function pertaining to the x-rescaling model  with
the nucleon off-mass shell, i.e. $F_2^{N/A}(x_{A},Q^2,k_1^2)
\rightarrow F_2^{N}(x_{A},Q^2) =F_2^{N} (\frac{
x_{Bj}}{z_1^A},Q^2)$,
 where   $z_1^A=k_1\cdot q/(m_p\nu)$ with $k_1^0 =M_A-\sqrt{{M_{A-1}^*}^2+{{\bf k}_1}^2}$;
\item
the structure function from Ref.~\cite{CKFS},  which assumes that
 the reduction of nucleon point like configurations (PLC)  in  medium (see Ref.\cite{fs}) depends upon  the nucleon virtuality:
\begin{equation}
F_2^{N/A}(x_{A},Q^2,k_1^2) \rightarrow F_2^{N/A}
 \left ( { x_{Bj}/z_1^N, Q^2}  \right ) \delta_A(x_{Bj},v(|{\bf k_1}|,E)),
\label{del}
 \end{equation}
  where
$z_1^N=(m_N +|{\bf P}_{A-1}|\cos\theta_{A-1})/m_N$. Here the reduction of PLC is given by the
quantity  $\delta_A(x_{Bj},v({\bf k},E))$, which depends upon the nucleon virtuality
(see \cite{CKFS}):
\begin{eqnarray}
\hspace{-0.5cm} v(|{\bf k}_1|, E)=
\left(M_A -\sqrt{(M_{A}- m_N+E)^{2}+{\bf k}_1^{2}}\right)^2-{\bf k}_1^2
- m_N^2.
\label{virtuality}
\end{eqnarray}
\end{enumerate}
\noindent where $E$ is the nucleon removal energy. It should be stressed that the two medium-dependent structure
functions provide similar results for the inclusive cross section
and that our aim is to answer the question as to
whether the  SIDIS experiment we are proposing could discriminate
between the two models. The results of calculations corresponding
to the kinematics $E_e = 12\,\,GeV$,  $Q^2\,\,=8\,\,{(GeV/c)}^2$,
 $\theta_{A-1} =145^{o}$,  $x_{Bj} = 0.45 $, $x_{Bj}^{\prime}=
0.35$, are presented in Fig.~\ref{Fig7}.
It can be seen that the discrimination between different models of the virtuality dependence of $F_2^{N/A}(x_{A},Q^2,k_1^2)$
can indeed be achieved by a measurement of the ratio~(\ref{ratioa}); as a matter of fact at $|{\bf P}_{A-1}| \simeq 0.4\,\, GeV/c$
the two structure functions differ by about $40\, \%$.

\section{Summary and conclusions}

We have considered the SIDIS process $A(e,e'(A-1))X$ on complex nuclei proposed in Ref.~\cite{scopetta} within the spectator
model and the PWIA,  and extended in Ref.~\cite{ciokop} by the inclusion of the FSI between the hadronizing debris and the
nucleons of the detected nucleus $(A-1)$.
The results of our calculations for the process $^2H(e,e'p)X$ show that the experimental data  can be well  reproduced in
the kinematics when the proton is emitted mainly backward in the range $70^{0} \lesssim \theta_{\bf p} \lesssim 145^{0}$,
 with the effects of  FSI  being very small in the very backward  direction, and dominating the cross section
 around $\theta_{\bf p} \simeq 90^{0}$. It is very gratifying to see that the experimental  data can be reproduced in a
 wide kinematical region, which makes us confident of the correctness of the spectator model and the treatment of the FSI
  between the nucleon debris and the detected proton. At emission angles $\theta_{\bf p} \lesssim 70^{0}$, the number of
   detected proton is much higher than our predictions, which is clear evidence of the presence of production mechanisms
    different from the spectator one. Among possible mechanisms  leading to forward proton production, target and/or current
    fragmentation should be the first process to be taken into account.
The first one has been analyzed in Ref. \cite{veronica}, finding that it contributes only forward and at proton momentum values
much higher than the ones typical of the Jlab kinematics. The contribution from current fragmentation
effects is under investigation.
The process on  nuclei with $A>2$ would be extremely interesting, since  the only mechanism for producing a recoiling
 $(A-1)$ nucleus would be the spectator mechanism.
These experiments, as stressed in Ref. \cite{ciokop}, would be very helpful to clarify the mechanism of the early stage
 of hadronization at short formation times without being affected by cascading processes, unlike the DIS inclusive hadron
 production $A(e,e'h)X$ where most hadrons with small momentum originate from cascading of more energetic  particles.
We have illustrated how, by measuring the reduced cross section on two different nuclei at the same value of the detected
momentum, the validity of the spectator mechanism and information on the survival probabilities of the spectator nuclei,
i.e. on the hadronization mechanism, could be obtained; moreover,  by measuring the cross section on the same nucleus,
but at two different values of $x_{Bj}$, the EMC effect could be tagged.
Experimental investigation of the processes $^2H(e,e'p)X$, $^3He(e,e'\,d)X$,  $^4He(e,e'\ ^3H)X$ and $^4He(e,e'\ ^3He)X$ would be extremely interesting and useful, and it is gratifying to see that they are planned to be performed thanks to the developments of proper recoil detectors~\cite{proposals}.

\section*{Acknowledgments}
 We would like to thank Sebastian Kuhn for providing the experimental data that we used to produce Fig.~\ref{Fig3} and
 for many  useful discussions and suggestions. Discussions with Alberto Accardi, Kawtar  Afidi, Boris  Kopeliovich and
  Veronica Palli are also gratefully  acknowledged. Calculations have been performed at CASPUR facilities under the
  Standard HPC 2010 grant \textit{SRCnuc}. One of us (L.P.K) thanks the Italian Ministry of Education,  University
  and Research (MIUR), and  the Department of Physics, University of Perugia, for support through the
   program ''Rientro dei Cervelli''.

\end{document}